# Shadow-Free Scattering with Complex Frequency Excitations

Vitali Kozlov[1], Wei Wang[1,2], Seunghwi Kim[1], Arno Thielens[1], and Andrea Alù[1,2]

[1]*Photonics Initiative, Advanced Science Research Center, City University of New York, New York, New York 10031, USA*
[2]*Physics Program, Graduate Center, City University of New York, New York, NY 10016, USA*

***Abstract:*** *The optical theorem dictates that the wave scattering of an object in the forward direction is proportional to the total power extinguished from the excitation. For passive objects, this relationship is a by-product of the intuitive notion that scattering and absorption necessarily create shadows, and that zero forward scattering, i.e., no shadow, can only be obtained if the object is fully transparent, i.e., it neither scatters nor absorbs. Materials with gain can be in principle leveraged to suppress forward scattering while preserving a large scattering signature in other directions. However, sufficient optical gain is challenging to realize, and it inherently leads to instabilities and nonlinearities. Here, we experimentally demonstrate shadow-free \ scattering from a passive high-index dielectric sphere by exciting it with a complex frequency signal tailored to mimic the presence of material gain. We achieve as much as 26dB suppression of the forward scattering when compared with optimal monochromatic excitations, while preserving strong scattering in other directions. More broadly, our results demonstrate how complex frequencies offer a natural pathway to enhance control over the scattering signature of passive objects, with implications for a range of photonic applications.*

## Introduction

One of the earliest demonstrations of the wave-like properties of light was the Arago experiment[1], which settled the debate on whether light behaves as a particle or a wave and confirmed how wave-based diffraction phenomena at the edge of an illuminated opaque disk can generate a bright spot at the center of its shadow. Many decades later, Kerker et al.[2] studied the scattering from magneto-dielectric spheres in the long-wavelength limit, proposing that, under suitable conditions on the material properties, it may be possible to achieve perfect scattering suppression in either the backward or the forward direction. While the zero-backward scattering condition has been pursued extensively by the metamaterials community to realize directional wave scattering, with impact in diverse fields such as antenna[3] and nanoantenna design[4], flat optics[5] and metasurfaces[7–11], the zero-forward scattering condition has faced more challenges:[12–14] a zero-forward scattering object that still absorbs and/or scatters in other directions appears to violate the optical theorem, which requires the extinction cross section of a linear object to vanish when the imaginary part of the forward scattering amplitude is zero[15]. For passive materials, this condition in turn requires that zero forward scattering is possible only for transparent scatterers, which do not scatter nor absorb at all, in contrast with Kerker's prediction. The apparent paradox of zero-forward scattering particles was theoretically resolved in a full-wave analysis of the scattering problem, including radiative corrections[16,17]. The resulting conclusion is that forward zero scattering is impossible for passive particles at any fixed frequency[18,19], and a minimum of forward scattering can be achieved at the Kerker condition only for deeply subwavelength particles that scatter very little[16].

A natural way to achieve a zero forward scattering particle that still complies with the optical theorem is to exploit active materials. Material gain implies that a particle can amplify incoming waves as they scatter off, enabling a negative absorption cross-section that can compensate for scattering in the power balance at the basis of the optical theorem, supporting the possibility of a zero net extinction, hence shadow-free scattering. Yet, the response requires a delicate balance of radiative loss and amplification, which may be hard to achieve with realistic material gain values without incurring in nonlinearities and instabilities.

Recently, it was theoretically suggested that zero forward scattering may be achieved by exciting a passive particle with temporally shaped waveforms that oscillate at complex-valued frequencies, rather than with continuous waves[20]. The field of complex frequency excitations has emerged in recent years as a solution to loss compensation in numerous photonic subfields and applications[21,22]. Indeed, the ability to control propagation[22], improve molecular sensing[23] and improve imaging resolution[24–28] has brought complex frequency excitations to the spotlight. Owing to the fact that these excitations can mimic material gain in a controllable manner implies that they may be in principle used to enable zero forward scattering. In the following, we theoretically explore the fundamental mechanisms as to how a passive object can leave no shadow while scattering at other angles when excited by such a temporally-tailored signal, and experimentally demonstrate this phenomenon at microwave frequencies.

## Zero-forward scattering at complex-valued frequencies

We start by analyzing the implications of the optical theorem for complex-valued frequencies. While time-domain versions of the optical theorem exists[29–31], it is most appropriate here to analytically continue in the complex frequency plane the frequency-domain optical theorem:[32]

$$P_a(s) + P_s(s) = \frac{\pi}{s\mu}\widehat{\boldsymbol{E}}_i \cdot [e_i(s)\overline{\boldsymbol{F}}_{\theta=0}(-s) - e_i(-s)\overline{\boldsymbol{F}}_{\theta=0}(s)], \quad (1)$$

where $P_a(s)$ and $P_s(s)$ are the absorbed and scattered powers, $s = \omega_i - i\omega_r$ is a complex-valued frequency, where $\omega_r = 2\pi f_r$ and $\omega_i = 2\pi f_i$ are the real and imaginary components of the operation angular frequency, respectively (see Supplementary Note 1 on conventions), $\mu$ is the free space permeability, $e_i(s)$ is the Laplace transform of the incident electric field, $\widehat{\boldsymbol{E}}_i$ is the incident polarization, and $\overline{\boldsymbol{F}}_{\theta=0}(s)$ is the scattering amplitude in the forward direction (behind the scatterer). The usual optical theorem is obtained by replacing $s = -i\omega_r$ in Eq. (1) and employing conjugate symmetry.

Consider a dielectric sphere of radius $a$ and relative permittivity $\varepsilon_r$, illuminated by a linearly polarized plane wave with electric field $\bar{E}_i(s,z) = e_i(s)e^{-is\frac{z}{c}}\hat{x}$. The scattered far-field is proportional to the scattering amplitude $\bar{E}_s \approx \frac{e^{-s\frac{r}{c}}}{r}\bar{F}(\theta,\varphi,s)$, which can be written to first-order in the Mie expansion[33] (see Supplementary Note 2 for details) as

$$\begin{aligned}\overline{\boldsymbol{F}}(\theta,\varphi,s) = \overline{\boldsymbol{M}}(\theta,\varphi,s)e_i(s) &\approx \overline{\boldsymbol{M}}_1(\theta,\varphi,s)e_i(s) \\ &= \frac{3c}{2s}\big[\big(a_1(s) + \cos(\theta)b_1(s)\big)sin(\varphi)\widehat{\boldsymbol{\varphi}} \\ &- \big(cos(\theta)a_1(s) + b_1(s)\big)cos(\varphi)\widehat{\boldsymbol{\theta}}\big]e_i(s),\end{aligned} \quad (2)$$

where $\overline{\boldsymbol{M}}(\theta,\varphi,s)$ is the linear Mie transfer function in the complex-frequency domain, connecting the incident and scattered fields, $\theta$ and $\varphi$ are the polar and azimuthal angles in the scattering plane ($\theta = 0$ is the forward direction in this convention), and $c$ is the speed of light. $a_1(s)$ and $b_1(s)$ are the first-order Mie coefficients, determining the strength of the radiated electric and magnetic dipole moments, respectively. While this solution is typically derived for real-valued frequencies, it can be analytically extended to the complex frequency domain. In the plane orthogonal to the incident polarization ($\varphi = \frac{\pi}{2}$), the transfer function takes the simple form $\bar{\boldsymbol{M}}_1\left(\theta, \frac{\pi}{2}, s\right) \propto [a_1(s) + b_1(s)cos(\theta)]\hat{\boldsymbol{x}}$, which reveals Kerker's condition for zero forward ($\theta = 0$) scattering[2] $a_1(s) = -b_1(s)$. It is worth noting that this condition remains true for higher orders in the series as well[34], i.e., as long as we can make sure that each electric Mie moment is equal and opposite to the corresponding magnetic moment, the forward scattering is bound to be zero.

As discussed in Ref. 16, for real-valued frequencies this condition cannot be satisfied with passive materials. This can be understood already by noticing that passivity requires the absorbed and scattered powers in Eq. (1) to be non-negative, hence identically zero forward-scattering must imply zero absorbed and scattered power. Hence, forward scattering suppression is only possible in the trivial case of a fully transparent object for any real-valued excitation frequency. A by-product of this challenge is that the Kerker condition $a_1(s) = -b_1(s)$ cannot be met at real frequencies. By substituting the scattered field from Eq. (2) into Eq. (1) and solving for the Mie coefficients we find in the lossless scenario (see Supplementary Note 2.2 for a full derivation):

$$a_1(s)a_1(-s) + b_1(s)b_1(-s) = \frac{1}{2}[a_1(s) + a_1(-s) + b_1(s) + b_1(-s)]. \tag{3}$$

For real-valued frequencies, this equation dictates that $Re\{a_1\} = |a_1|^2$ and $Re\{b_1\} = |b_1|^2$ for a lossless sphere[16], and generally holds true for all other higher-order Mie scattering coefficients. Given that the real parts of $a_1$ and $b_1$ must be non-negative, the only way to satisfy $a_1 = -b_1$ is for them to be zero, which again yields zero total scattering. Minimum forward scattering can be achieved when the Kerker condition is satisfied for the imaginary parts of the Mie coefficients, while the real part is sufficiently small, but both real parts must be necessarily positive and interfere constructively in the forward direction. As we show in the following, for complex-valued frequencies, these constraints are significantly relaxed, and it is indeed possible to find complex-valued frequencies in which $P_a(s) = -P_s(s)$ and $a_1(s) = -b_1(s)$ while complying with causality and passivity[20].

This condition is illustrated in Fig. 1(a), in which the same sphere is illuminated by real-valued and complex-valued frequency waves. The real-frequency waves must leave an ordinary shadow in the forward direction, while the complex frequency wave may yield zero forward scattering, leaving a shadow-free spot. Quite remarkably, the required complex-valued frequency excitation that achieves this feat can be evaluated by studying in the complex frequency plane the forward scattering Mie transfer function, derived from Eq. (2) in Supplementary Section 2, and whose magnitude is shown in Fig. 1(b) for a sphere with $\varepsilon_r = 19.74$, $a = 12.04mm$. While on the real frequency axis this function must be non zero, in the lower half of the complex frequency plane it can support poles and zeros (marked by purple dashed circles). A negative imaginary part of the excitation frequency ensures decaying fields, for which virtual gain emerges[21]. In essence, at these complex frequencies, the excitation fields engage the scatterer resonances, which store and delay the excitation in a timed way, such that the scattered field at any instant in time interferes destructively with the leaked stored energy in the forward direction. Quite remarkably, when excited with the specific waveform that targets this complex-valued zero, after a transient we can expect the system to reach a quasi-steady state response in which the forward scattering is suppressed. Figs. 1(c-f) show the calculated Mie coefficients (real and imaginary parts) for the first complex-valued zero shown in Fig. 1b and for the real frequency obtained by setting the imaginary part to zero. As expected, for real frequencies (Fig. 1c), the real part of the Mie coefficients are always positive, preventing them from obeying the desired condition $a_1 = -b_1$. The imaginary parts, however, are of opposite sign at the design frequency (Fig. 1e, dashed pink line). If we analyze the same coefficients at the complex frequency supporting a zero of the forward scattering transfer function (Figs. 1d,f), we do find that the real part of the scattering coefficients can now yield opposite signs, and destructively interfere in the forward direction. The resulting scattering patterns, shown in Figs. 1(g,h) do yield remarkably different responses, and the expected zero forward scattering for the complex frequency excitation.

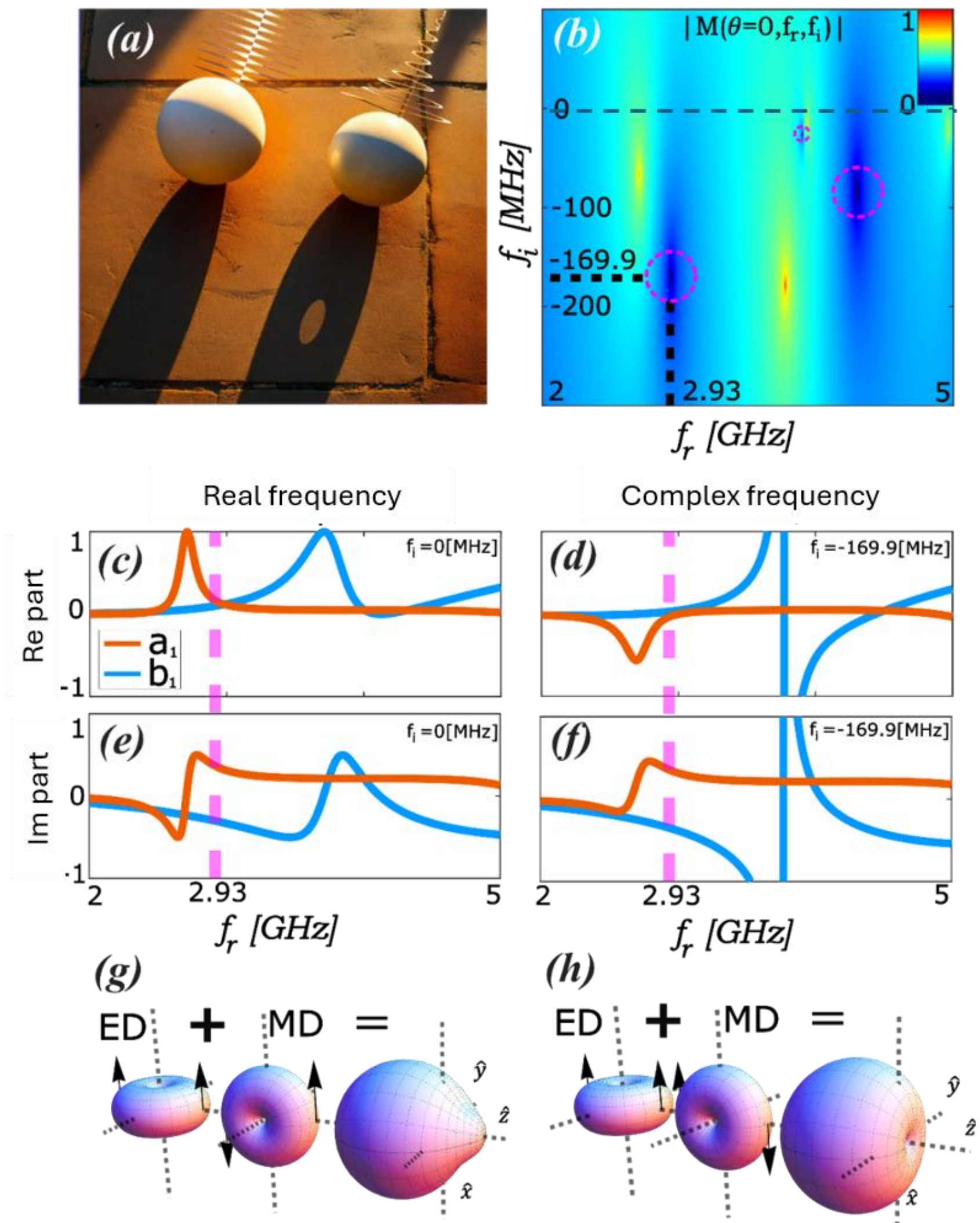

Fig.1| **(a)** Illustration of zero forward scattering and shadow-free response for complex frequency excitations. The left sphere is illuminated by a real frequency wave, and in compliance with the optical theorem supports a shadow proportional to the total scattered and absorbed power. The right sphere is excited with a complex frequency wave supporting zero forward scattering, fully transmitting the incident signal at the back of the scatterer. **(b)** Magnitude of the $\hat{x}$-polarized forward Mie transfer function analytically continued in the complex frequency plane. Zeros are marked by purple circles, and the lowest-order zero is located at $f_r + if_i$ = 2.93 – 0.17i [GHz]. **(c)-(f)** Real and imaginary parts of the first two Mie coefficients, associated with the electric (ED) and magnetic dipole (MD) scattering. Here, **(c)** and **(e)** are calculated for real-frequency excitation ($f_i = 0$), while **(d)** and **(f)** are for complex-frequency excitation ($f_i = -170MHz$), targeting the zero of the transfer function. **(g)-(h)** Resulting scattering patterns from the superposition of ED and MD contributions, revealing zero forward scattering for the complex frequency excitation.

The realization of zero forward scattering for tailored complex frequency excitations is consistent with the concept of virtual gain for temporally decaying excitations of a resonant system. As discussed in Supplementary Note 3, the measured forward scattering cannot yield an exactly zero value, since causality enforces some conditions on the Mie transfer function associated with the measurement bandwidth of the system. Yet, we can expect a drastically reduced scattering magnitude in the forward direction by exciting at the proper frequency. Any practical realization will observe such drastic reduction of forward scattering after a transient, as the scattered fields start decaying at the complex frequency of the excitation in the quasi-steady state [21,35], provided that no slower decaying poles exist in the proximity of the forward scattering zero (see Supplementary Notes 4, 5). We study these phenomena experimentally in the following.

## Experimental observation of zero-forward scattering

The results in Fig. 1 are limited to a complex frequency analysis based on analytical continuation of the Mie scattering boundary value problem, hence they do not consider the transient response of the system and assume a quasi-steady state condition in which the scattered fields oscillate at the same complex frequency as the excitation[21]. In order to validate our predictions, we carried out an experiment for the same geometry studied in Fig. 1, excited at microwave frequencies. We selected a high index dielectric sphere in order to minimize absorption and to enable a sufficiently strong resonant response within the subwavelength frequency range in which ED and MD are dominant, such that the forward scattering amplitude zeros lie close to the real axis, ensuring a slow decay of the excitation fields. This is consistent with the physics of the problem: since the particle obeys passivity, the cancellation of the forward fields must arise through destructive interference of the scattered fields from the excitation and the leakage of stored energy from the resonance. As such, a high-Q resonance ensures higher stored energy, and hence a lesser requirement on the excitation decay.

The scatterer was placed at the center of the rotating stage of an anechoic chamber (Fig. 2a), while a transmitting linearly polarized horn antenna and a receiving linearly polarized rectangular probe antenna located in the far field were used to characterize the scattering. Both antennas, with an operational bandwidth between 2 and 5GHz, were orthogonally polarized to the rotational plane, which is the only relevant polarization due to the scatterer symmetry. The stage with the sphere at its center was rotated between -130° to 130° with 10° steps, with the antennas placed in the far-field region. Scattering-parameter measurements were retrieved with and without scatterer for calibration (see Supplementary Note 6).

Figure 2(b) shows the scattering amplitude as a function of angle and real frequency calculated using the Mie expansion in Eq. (2). The corresponding experimental results in Fig. 2c match well with these predictions, validating our model (see Supplementary Note 6.2). Three vertical black dashed lines in Fig. 2(b) and 2(c) indicate the real frequency for which the forward scattering supports zeros in the complex frequency plane, in accordance with Fig. 1(b). The purple dotted vertical lines mark the boundaries of the effective real-frequency bandwidth associated with the complex-frequency excitation, calculated by evaluating the Fourier spectrum associated with the complex frequency of each scattering zero. As expected, a narrower bandwidth corresponds to a zero located closer to the real axis.

Fig. 2(d) shows the experimentally extracted values of the real part of the dipolar Mie coefficients, which fit very well with the numerical values shown in Fig. 1(f). We evaluated them (see Supplementary Note 6.3) by synthesizing the corresponding complex frequency excitation by combining in Fourier space the real-frequency measurements in Fig. 2(c), and normalizing to the excitation amplitude. Remarkably, we do observe the real part of the ED coefficient to become negative around the complex frequency zero, in apparent contradiction with the passivity requirement, validating the emergence of virtual gain at complex frequencies. An alternative way of validating this result is to consider the real part of the reciprocal of the Mie coefficients $Re[1/a_1(s)], Re[1/b_1(s)$, shown in Fig. 2(e). For a passive scatterer with negligible absorption these values are equal to 1 for all real frequencies[36], marked with a black dashed line, and in the presence of absorption their value is above 1. Our retrieved real parts of the reciprocal Mie coefficients under complex frequency excitations, on the contrary, dive below unity, indicating the emergence of effective gain in the scattering problem.

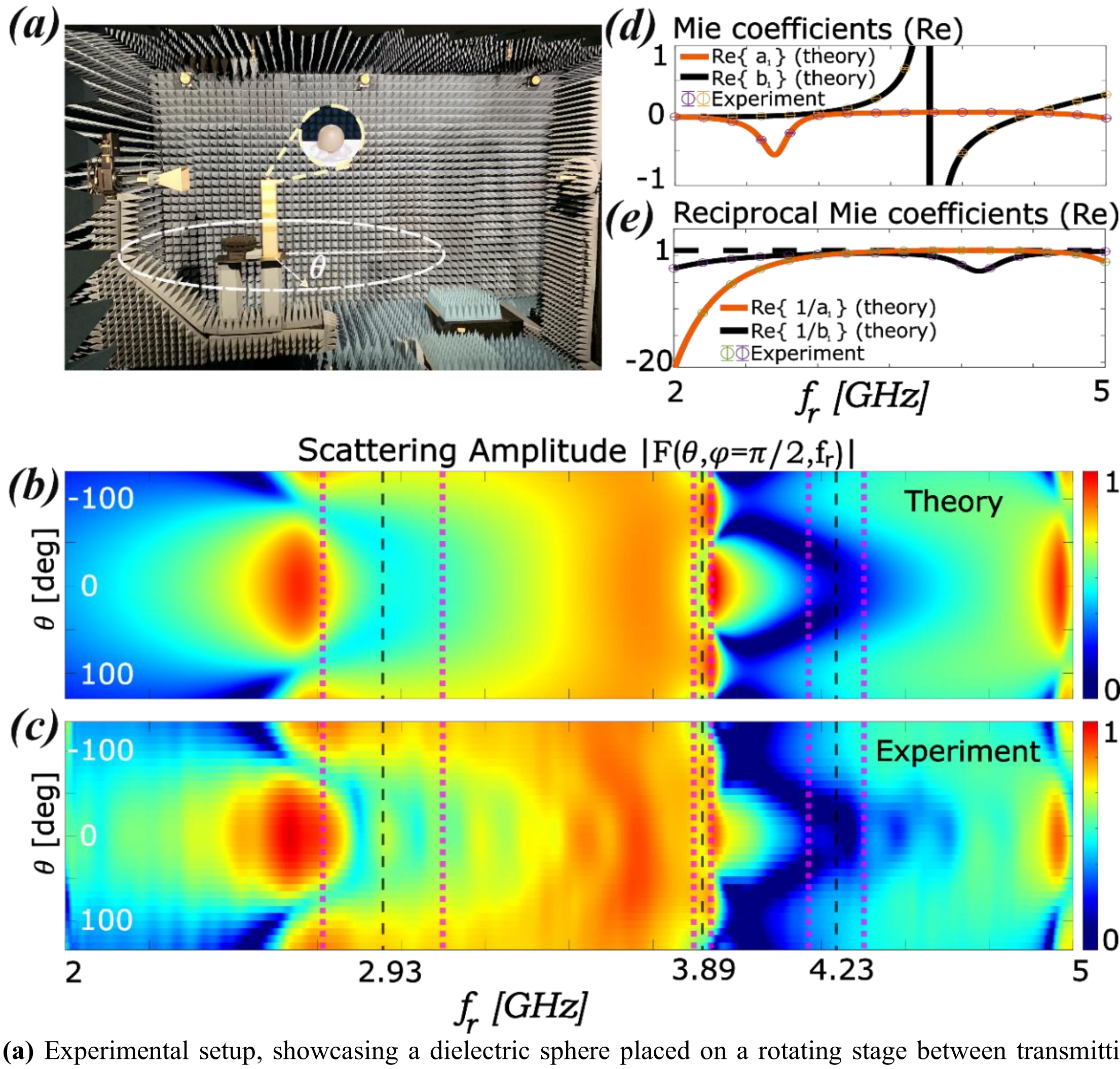

Fig.2| **(a)** Experimental setup, showcasing a dielectric sphere placed on a rotating stage between transmitting and receiving antennas. **(b)** Calculated scattering amplitude versus angle and real frequency. The dashed black lines indicate the real parts of the forward scattering zeros, while dotted purple lines delimit the boundaries of the effective bandwidth for $f_r \pm f_i$ **(c)** Experimental measurement of the scattering amplitude. **(d)** Experimental versus theoretical calculation of the real part of the retrived Mie coefficients. **(e)** Experimental versus theoretical calculation of the real part of the reciprocal Mie coefficients.

Having established that our scatterer supports zero forward scattering at complex frequencies, we explore its temporal dynamics for a realistic excitation and in particular the transient regime that leads to a quasi-steady state. Rather than exciting the scatterer with a complex wave form, by exploiting the linearity of the problem, we synthesize a realistic complex frequency excitation by combining in amplitude and phase a range of real-frequency measurements. This strategy was recently employed in other complex frequency excitation setups, as well as to explore other time-domain responses[22,23,27,37]. Fig. 3(a) shows for a few selected scattering angles the time-domain response for excitation of the dominant forward scattering zero, located at the complex frequency $2.93[GHz] - 119.6i\ [MHz]$. The dense carrier oscillations in these plots are removed using digital down conversion, leaving the envelope alone (see Supplementary Note 7).

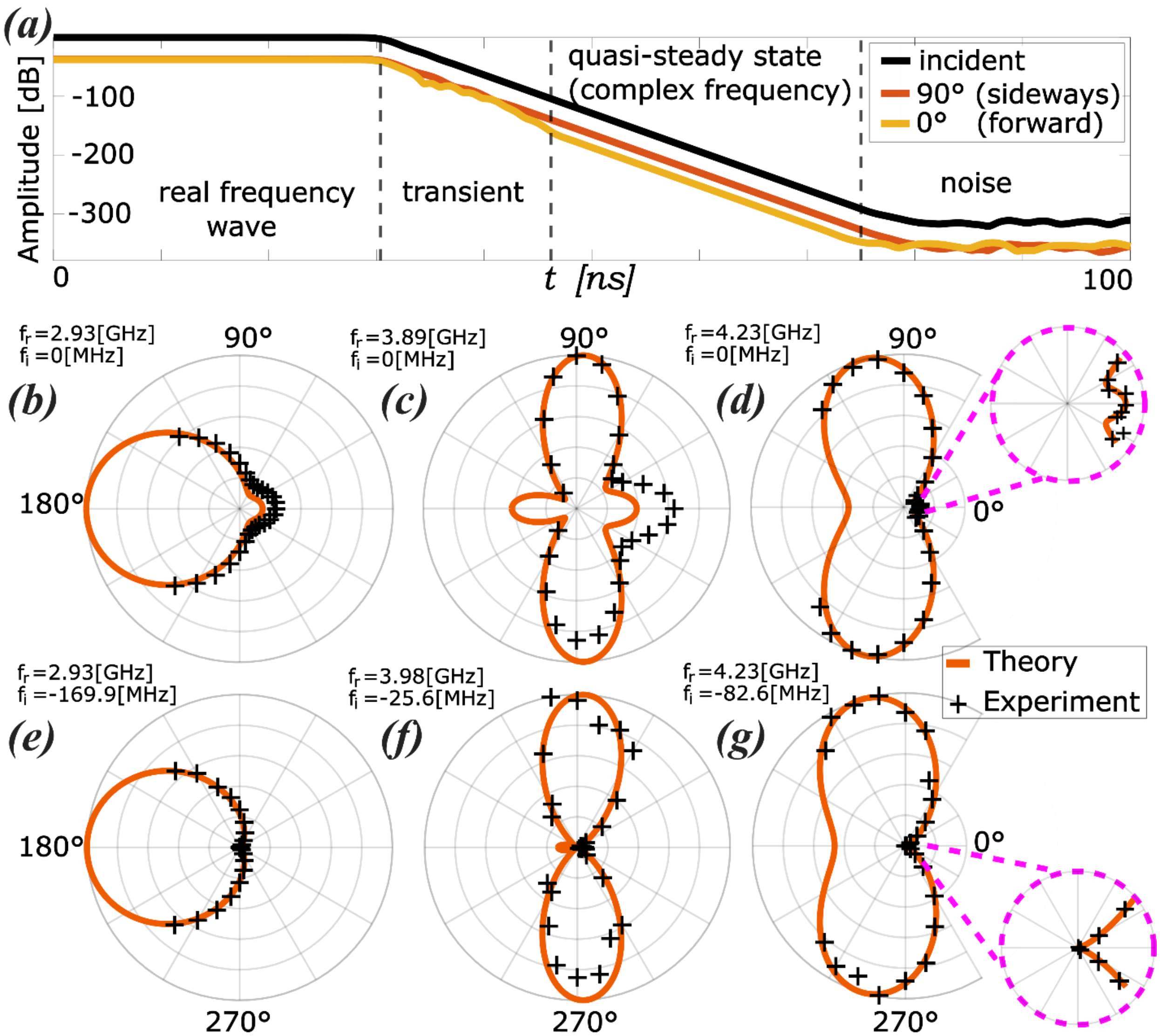

Fig. 3| **(a)** Scattered and incident field temporal envelopes as a function of time. Different dynamic regions (real frequency wave, transient, quasi-steady state and noise) are shown in response to an incident excitation that undergoes switching from purely real $2.93[GHz]$ to a complex frequency wave of $2.93[GHz] - 119.6i\ [MHz]$, tailored to excite the forward-scattering zero. Resulting normalized scattering patterns as a function of $\theta$ for complex and real frequencies at $f_r + if_i$ = **(b)** $2.93[GHz]$ **(c)** $3.89[GHz]$ **(d)** $4.23[GHz]$ **(e)** $2.93[GHz] - 119.6i\ [MHz]$ **(f)** $3.89[GHz] - 25.6i\ [MHz]$ and **(g)** $4.23[GHz] - 82.6i\ [MHz]$ ,where the purple dashed insets zoom in on the scattering in the forward direction, showing strong forward suppression under complex frequency wave excitation.

Figure 3(a) shows the scattered field envelopes for an incident excitation switching from a real frequency wave to a complex frequency harmonic with same real frequency. The scattered field shows a distinct transient period before reaching the quasi-steady state, and finally decaying below the noise floor (see Supplementary Note 8). Three envelopes are shown, corresponding to the incident excitation, forward scattered field and right-angle (measured at 90°) scattered field. While during the real frequency excitation the forward-backward envelope ratio does not exceed 3.7dB, a contrast of 22dB is achieved in the quasi-steady state under the optimal complex frequency excitation, as shown in Fig. 3(a). By repeating the synthetic temporal processing for all three forward-scattering zeros in the complex frequency plane, the scattering patterns were calculated and plotted in Fig. 3(b-g). These values were measured by taking the square of the scattered field amplitude at each angle at the same moment in time during the quasi-steady state, and normalizing to the strongest scattering direction.

Figures 3(b-d) show the scattering patterns for real frequency waves while Figs. 3(e-g) display the patterns for complex frequency excitations. Clear forward scattering suppression is observed for all three complex frequency excitations, in good agreement with theoretical predictions, which were obtained from the analytical continuation in Eq. (2). The most pronounced suppression is achieved for the third zero in Fig. 1(b), as seen in Fig. 3(g), where the value of the forward scattering direction is -72dB, which is a suppression of 26dB compared to the forward scattering at real frequencies shown in Fig. 3(e). Finally, it is interesting to note that Fig. 3(f) displays strong suppression in both the forward as well as backward directions compared to the real frequency excitation in Fig. 3(c). While the two directions have Mie transfer function zeros that are located close by, they are not exactly at the same location in the complex frequency plane, preventing fully symmetric suppression [20,33,38]. Equation (3) and the generalized Mie coefficient constraints can be thought of as the physical origin of virtual gain in the lossless sphere, allowing the real part of the coefficients to decrease below zero, just like spheres with active media.

## Conclusions

In this work, we have investigated virtual gain associated with complex frequency excitations in scattering problems, shedding light to the exotic phenomenon of shadow-free scattering from passive objects. The use of complex-frequency optical theorem relaxes the limitations imposed on the forward scattering amplitude by its real-frequency counterpart, and enables Mie coefficients with negative real parts, which we predicted analytically and retrieved from experiments. While causality prevents achieving an identical zero forward scattering in the quasi-steady state due to restrictions on the finite bandwidth of any realistic excitation, it can approach arbitrarily small values. We demonstrated the phenomenon experimentally by employing complex frequency excitations in the radio-frequency range, observing over 2 orders of magnitude suppression of the forward scattering amplitude. While it is remarkable that a practical excitation of the scatterer with a realistic signal that oscillates at the complex-valued frequency retrieved from analytically continuing the Mie scattering boundary value problem, we stress that not all zeros can be accessed by a realistic excitation, due to limitations on the quasi-steady state condition[21]. Successful demonstrations require the poles near the real frequency axis to be sufficiently far away from the desired complex frequencies to avoid their excitations. Indeed, our experiments show best results when targeting the third zero in Fig. 1(b), which is further away from any pole. Beyond the demonstration of minimal forward scattering under complex frequency excitations, our work opens new avenues for wave control across a wide range of systems. These phenomena can be applied to beam steering in flat optics, directional light sources, light trapping for energy harvesting, efficient wireless power transfer, scattering control and noninvasive sensors. They may also inspire advances in electrically small antennas with high performance and enhance light-matter interactions for quantum applications[43,44]. Furthermore, our technique demonstrates a nontrivial scattering phenomenon that leverages virtual gain at complex frequencies, suggesting opportunities in loss compensation in a variety of resonant phenomena, which can be extended by integrating concepts from non-Hermitian physics[39,40].

## Acknowledgements

This work was supported by the Department of Defense, the Simons Foundation, the European Research Council and the National Science Foundation.